\documentstyle[12pt,preprint,aps,prl]{revtex}     

\tighten
\draft

\begin{document}

\preprint{NCU-CCS-980905}

\title{An accelerated charge is also absorbing power%
}

\author{James M. Nester\thanks{email: nester@joule.phy.ncu.edu.tw}
\and C. M. Ko\thanks{email: cmko@joule.phy.ncu.edu.tw}
}

\address{Department of Physics and Center for Complex Systems,
National Central University,\\ 
 Chungli, Taiwan 320, R.O.C.}

\date{\today}

\maketitle

\begin{abstract}
An accelerated classical point charge radiates at the Larmor power
rate $2e^2a^2/3$, leading to the expectation of an associated
radiation reaction force.  The famous Abraham-Lorentz-Dirac proposal
is plagued with difficulties.  Here we note a simple, amazing, and
apparently overlooked fact:  an accelerated charge is also always
absorbing power at exactly the Larmor rate.  Implications for
radiation reaction and the particle motion are considered.  Our
analysis supports Kijowski's recent proposal.
\end{abstract}

\pacs{PACS number(s): 41.60.-m, 03.50.De}



Particles and fields are central concepts in modern
physics.  The idea of particles interacting through fields is
a key paradigm.  Its roots are classical, primarily in
electrodynamics where Faraday's field idea replaced Newton's
``action at a distance''.  Classical point charges
interact locally with the electromagnetic field via the Lorentz force
law $\dot{\bf p}={\bf F}=e({\bf E}+{\bf v}\times {\bf B})$.  This
relation, essentially the genesis of the field idea, both {\em
defines\/} the electromagnetic field and {\em predicts} the particle
motion---with the exception of {\em self interaction\/} effects.

From Maxwell's field equations it was found that an accelerating
classical point charge radiates power at the Larmor rate.  Since the
particle apparently looses energy, one expects an associated radiation
reaction (damping, friction) force.  For over a century many famous
physicists struggled with this idea and various associated models for
classical particles.  The best known reaction force proposal was
developed, based on an extended charge model, by Lorentz.  Later Dirac
gave a derivation of the relativistic version for point particles
using the conservation laws.  However, the equation of motion, with
this radiation reaction force included, has been plagued with numerous
difficulties.  Here we note a very relevant, simple, amazing fact,
which has apparently been overlooked for over 100 years:  an
accelerated point charge is also {\em always absorbing power} at {\em
exactly} the Larmor rate.  Implications for radiation reaction and the
particle motion are considered.  Our analysis supports Kijowski's
``renormalized electrodynamics''.

Maxwell's equations \cite{E&M,Jac75,Roh65}
(we take $c=1$), $\partial_\mu F^{\nu\mu}=4\pi J^\nu$
with $F_{\mu\nu}=\partial_\mu A_\nu-\partial_\nu A_\mu$,
in the Lorenz gauge   $\partial_\mu A^\mu=0$,
lead to the wave equation $\partial_\mu
\partial^\mu A^\nu=-4\pi J^\nu$ for the potentials.
A solution can be written in terms of the potentials obtained via
the retarded or advanced Green function along with an associated
solution to the homogeneous equation:
$A^\mu=A^\mu_{\mbox{\footnotesize{in}}}+A^\mu_{\mbox{\footnotesize{ret}}}
=A^\mu_{\mbox{\footnotesize{out}}}+A^\mu_{\mbox{\footnotesize{adv}}}$,
whence
$F^{\mu\nu}=F^{\mu\nu}_{\mbox{\footnotesize{in}}}
+F^{\mu\nu}_{\mbox{\footnotesize{ret}}}
=F^{\mu\nu}_{\mbox{\footnotesize{out}}}
+F^{\mu\nu}_{\mbox{\footnotesize{adv}}}$.

A point charge $e$ moving along the spacetime path
$q^\alpha(\tau)$ generates a current,
 $J^\mu(x)=e \int_{-\infty}^{+\infty} v^\mu \delta^4(x-q(\tau))d\tau$,
where $v^\mu:=dq^\mu/d\tau$.
The resultant (Li\'enard-Wiechert) potentials are
$A^\alpha=e [v^\alpha/R]$,
where
 $R:=|R^\nu v_\nu|$, with $R^\nu:=x^\nu-q^\nu(\tau)$.
Here and elsewhere the brackets $[\quad]$ indicate that the enclosed
quantities are to be evaluated at the retarded/advanced time
determined implicitly by $R^\mu R_\mu=0$.  The associated
fields,
\begin{equation}
F^{\mu\nu}_{\mbox{\footnotesize{ret}}\atop\mbox{\footnotesize{adv}}}=
\pm \left[\frac{e}{R} \frac{d}{d\tau}
\left(\frac{v^\mu R^\nu - v^\nu R^\mu}{R} \right)\right],
\end{equation}
include two types of terms. One, proportional to $v^\mu$,
is of the Coulomb $\sim e/R^2$ form and is bound to the charge,
moving along with it. The other, proportional to the acceleration
$a^\mu:=dv^\mu/d\tau$, is long range, having $1/R$ fall
off; it is interpreted as radiation which escapes from the charge.
In the charge's instantaneous rest frame, in vector notation,
${\bf B}=[\hat{\bf R}\times {\bf E}]$, where
 \begin{equation}
 {\bf E}={\bf E}_{\mbox{\footnotesize{cou}}}+
{\bf E}_{\mbox{\footnotesize{acl}}}=
 e\left[ { \hat{\bf R} \over R^2} \right] +
 e\left[ { \hat{\bf R}\times ( \hat{\bf R}\times \dot{\bf v} ) \over R}
\right].
 \end{equation}
This splitting (based on the retarded/advanced decomposition) however,
although intuitively appealing, is really not physical, for
$\nabla\cdot{\bf E}_{\mbox{\footnotesize{acl}}}=
(2e/R^2){\bf a}\cdot{\hat{\bf R}}\ne0$
outside of the moving charge.
Hence ${\bf E}_{\mbox{\footnotesize{acl}}}$, taken by itself, appears
to have conjured out of the vacuum a rather strange source
charge density---the associated 4-current is proportional to $R^\mu$ and
is thus moving at the speed of light \cite{Tei71}.
However the ``acceleration'' fields really do, through the
$O(R^{-2})$ part of the Poynting vector,
 ${\bf S}_{\mbox{\footnotesize{acl}}}=
(1/4\pi){\bf E}_{\mbox{\footnotesize{acl}}}
\times{\bf B}_{\mbox{\footnotesize{acl}}}
=(1/4\pi)|{\bf E}_{\mbox{\footnotesize{acl}}}|^2\hat{\bf
R}=(e^2a^2/4\pi)R^{-2}\sin^2\theta\,\hat{\bf R}$,
dominate the power radiated out to large distances:
\begin{equation}
P=\oint {\bf S}_{\mbox{\footnotesize{acl}}}
\cdot \hat{\bf R}d\sigma={2\over3}e^2 a^2,
\end{equation}
the celebrated {\em Larmor power formula}.  Note however, that this
calculation is valid {\em only if} there are no other fields that
interfere with the radiation.  Real charges are actually immersed in a
sea of electromagnetic fields ranging from the $3$ K
cosmological microwave radiation thorough the output of the sun and
stars to radio, television and thermal radiation at $\sim 300$ K.

In practice one cannot measure the radiation at infinity.  Since the
radiation rate is proportional to the instantaneous acceleration of
the particle, it is inferred that any accelerating charge emits
radiation.  Rohrlich has rigorously established this as a {\em local
criterion} for radiation \cite{Roh61,Roh65}.  Hence our view is that
the emission of ``radiation'' is a local process happening near the
charge.  An accelerating charge causes a certain type of disturbance
of the electromagnetic field in its immediate neighborhood.  This
disturbance may propagate out to large distances.  Instead
it may suffer interference from other effects propagating in the
field, so that little or no power may actually get out to infinity;
but this is a vacuum field propagation issue; it should not be held to the
charged particle's account.  Note that, because of interference, an
outward propagating signal {\em does not} conserve power in general.

The usual dogma is that the emission of radiation is irreversible and
hence there is an associated radiation damping (or
friction).  If power is emitted then energy conservation
considerations lead to a radiation reaction force.  The orthodox
version is the Abraham-Lorentz-Dirac force
\cite{E&M,Jac75,Roh65,Dir38,WF45,Par87}.
 A simple argument (see, e.g., Jackson \cite{Jac75} \S 17.2) considers
the radiated energy
\begin{equation}
\int_{t_1}^{t_2} {2\over3}e^2\dot{\bf v} \cdot \dot{\bf v} dt
={2\over3}e^2 \left(  \dot{\bf v} \cdot{\bf v}{\Big|}_{t_1}^{t_2} -
 \int_{t_1}^{t_2} \ddot {\bf v} \cdot {\bf v} dt \right).
\end{equation}
Under certain conditions the first term on the rhs will vanish (e.g.,
if the motion is periodic or bounded) for a suitable choice of
interval.  Then we can identify the radiative reaction force:
\begin{equation}
{\bf F}_{\mbox{\footnotesize{rad}}}={2\over3}e^2\ddot {\bf v}.
\end{equation}
As Jackson says: ``It can be considered as an equation which includes
in some approximate and time averaged way the reactive effects of the
emission of radiation''.  But many regard this equation (or
its relativistic generalization) as fundamental (see e.g.,
\cite{Roh65,Tei71,Roh61,Dir38,Pla61,Com93}).
Wheeler and Feynman \cite{WF45} observed that: ``The existence of
this force of radiative reaction is well attested:  (a) by the
electrical potential required to drive a wireless antenna; (b) by the
loss of energy experienced by a charged particle which has been
deflected, and therefore accelerated, in its passage near an atomic
nucleus; and (c) by the cooling of a glowing body.'' But each of these
processes has its converse, for example a wireless receiver can absorb
power and a cool object can absorb heat radiation.  Microscopically,
the interactions of classical physics were presumed to be time
reversible.  If a charge could emit radiation it could likewise, under
appropriately reversed conditions, absorb radiation.  In classical
physics the asymmetry between past and future is only a macroscopic
statistical effect, certain processes are regarded as being much more
probable.  And so it should be with electrodynamics. Indeed Wheeler
and Feynman \cite{WF45} attest:  ``We have to conclude with
Einstein${}^{11}$ [citation in the original] that the irreversibility
of the emission process is a phenomenon of statistical mechanics
connected with the asymmetry of the initial conditions with respect to
time.'' Our usual physical intuition accepts an accelerated charge
emitting power, readily imagining outgoing waves carrying away energy,
but regards power absorption as uncommon, imagining it quite unlikely
that waves will be prearranged to focus on a charge.  We will show
that this picture of power absorption is inaccurate.

Although power absorption situations are actually physically common,
power absorption by accelerating charges is rarely mentioned
(except in connection with the direct interaction theory \cite{WF45}
where its role is equally as important as emission.).
However, the point we wish to make here is not just that a charge
could absorb power but that an accelerated charge {\em must}
constantly be absorbing at the same (Larmor) rate as it is emitting.

Including the aforementioned radiation damping force leads to the
famous Abraham-Lorentz-Dirac (ALD) equation for a moving point charge
\cite{Dir38,E&M,Jac75,Roh65}, which has been plagued with problems.
Briefly, the ALD equation
(i) is a 3rd order equation,
(ii) seems to violate time reversiblity,
(ii) has runaway solutions,
 (iii) has ``unphysical'', nonexistent,
nonunique and counter intuitive solutions,
(iv) violates causality (preacceleration---for an ingenious
resolution see \cite{Val88}).  Much research has been devoted to these
problems \cite{Par87}, occasionally the controversy has become heated
\cite{Com93}.  Dissatisfaction with the ALD has led to many ingenious
proposals (see, e.g.,  \cite{Par87,Yag92,HL95}).
In our opinion most of the discussions overlook, or at best do
not give enough weight to, certain fundamental principles of classical
physical interactions:

1. they are time reversible
(QED is also time reversible):
not only does every emission situation have a corresponding
absorption situation, microscopically they are on par and any
difference in rate is a consequence of a global asymmetric boundary
condition (as Wheeler, Feynman and Einstein believed)
\cite{timerev}.

2. they have 2 initial data per degree of freedom: by this
reckoning the ALD has $9/2$ degrees of freedom.

3. they are local and instantaneous: an influential
text states ``physics is simple when analyzed locally'' \cite{MTW}.

To establish our point about absorption,
let us first consider Born's solution \cite{Bor09,FR60,Roh65}
for a uniformly accelerating charge. The path is hyperbolic: $x=0$,
$y=0$, $z=(\alpha^2+t^2)^{1/2}$.  The field, in cylindrical
coordinates, is given by $E_\phi=B_\rho=B_z=0$, and
\begin{eqnarray}
\qquad B_\phi&=&8e\alpha^2\rho t/\xi^3,\qquad
E_\rho=8e\alpha^2\rho z/\xi^3, \nonumber\\
E_z  & =&-4e\alpha\left(\alpha^2+t^2 +\rho^2-z^2\right)/\xi^3,
\end{eqnarray}
with
$
\xi=\left\{4\alpha^2\rho^2+
(\alpha^2+t^2-\rho^2-z^2)^2\right\}^{1/2}
$,
where $\alpha=a^{-1}$.  Note that
$\bf E$ is time symmetric, and $\bf B$ is time antisymmetric.  Note
also that the solution has a boost Lorentz symmetry:  the field values
at a later proper time are just obtained by Lorentz transforming
between the respective instantaneous inertial rest frames.  Thus all
times are essentially equivalent.  Without loss of generality we can
examine the fields in the frame in which the charge is at rest at the
lab time $t=0$.

It has been noted that $\bf S$ vanishes at
$t=0$, since $\bf B$ vanishes at that instant.  Pauli \cite{Pau58}, in
particular, has interpreted this fact to mean that a uniformly
accelerated charge does not radiate.  However, it has been argued
(see, e.g., \cite{Roh65} \S 5.3) that for radiation we should look
along the null cone.  With $z=\alpha+\zeta$, calculating on the null
cone,
 $t^2=r^2:=\rho^2+\zeta^2$, of the point where the charge is
at rest at lab time $t=0$,  we find
$E_\rho=e\rho(1+a\zeta)/r^3$, $E_z=e(\zeta+a\zeta^2-ar^2)/r^3$,
$B_\phi=ea\rho t/r^3$.
Hence
 ${\bf S}\cdot \hat{\bf r}r^2=(t/4\pi r)e^2a^2\sin^2\theta $,
which yields, after integration over angles with $t=+r$,
the expected Larmor rate, $P=(2/3)e^2a^2/c^3$,
for the power radiated along the outgoing null cone (independent of r
in this case).  {\em But}, calculating the power {\em flowing into}
the charge along the {\em incoming} null cone of the same point,
$-t=r\ge0$, gives the negative of the standard Larmor rate!  The
emitted power is just the absorbed power.  Apparently power has come
from outside and has simply flowed through the location of the charge
to be reemitted.

Note that in \cite{Roh61} this solution was derived using only the
retarded interaction, so that all fields in the region $t+z>0$ should
have come from the charge.  But we have radiation incoming from the
past (tracing it back we find that it comes from outside the region
$t+z>0$) converging on the charge.  Where did the power come from?  We
observe that eternal acceleration is not physically realistic.  Power
flowing into a charge which began accelerating only one second ago
could not be traced far back along the incoming null cone (propagation
of power along the null cone is not conserved).

For uniform acceleration the power absorption exactly equals the
Larmor emission rate, which depends only on the acceleration.  Tracing
the radiation along the null cone suggests that even for a
non-uniformly accelerated particle the rates would still be equal and
would depend only on the instantaneous acceleration.  To establish
this result, we consider the the behavior of the instantaneous fields
near a charge undergoing a general acceleration.  In addition to a
regular part they include singular parts.  The electric field,
${\bf E}_{\mbox{\footnotesize{sing}}}={\bf E}_{-2}+{\bf E}_{-1}+{\bf E}_0$,
includes the unbounded for ${\bf r}\to0$ terms
\begin{equation}
{\bf E}_{-2}={e\over r^2}\hat{\bf r}, \qquad
{\bf E}_{-1}=-{e\over 2r} ({\bf a}+{\bf a}\cdot \hat{\bf r} \hat{\bf r}),
\end{equation}
as well as
${\bf E}_0=3e/8\{ ({\bf a}\cdot \hat{\bf r})^2\hat{\bf r} +
2{\bf a}\cdot \hat{\bf r}{\bf a} - a^2 \hat{\bf r}\}$.
The latter, like
${\bf B}_{\mbox{\footnotesize{sing}}}={\bf B}_0=
-(e/2)\dot {\bf a}\times \hat{\bf r}$,
is bounded but still singular: the
limiting value depends on the direction of approach for
${\bf r}\to0$.
(It is easily checked
that ${\bf B}_0$, ${\bf E}_0$ and ${\bf E}_{-1}$ have vanishing divergence
while $\nabla\cdot{\bf E}_{-2}=4\pi\delta^3({\bf r})$.)\ \
These expressions can be obtained from
the 4-covariant instantaneous expression (which played an important role in
Dirac's seminal paper, \cite{Dir38} eq.\ (60), and
Rohrlich's book, \cite{Roh65} eq.\ (6-68)) or by specializing the more
general expression given by Page way back in 1918 \cite{Page}, eqs.\
(23,24).  (It is amazing that so much of the significance of these
expressions was not appreciated until the recent work of Kijowski
\cite{Kij94}.)\ \ These singular {\it self field} values are in
general superposed with some regular solution to the homogeneous
equation.  The regular parts of the fields at the location of the
charge are just given by some constant values.  The particular form of
${\bf E}_{\mbox{\footnotesize{reg}}}(0)$ (and ${\bf
B}_{\mbox{\footnotesize{reg}}}(0)$) is determined by various
mathematical/physical choices like the Green function and boundary
conditions:  in particular using the retarded/advanced Green function
along with  ${\bf E}_{\mbox{\footnotesize{in/out}}}=0$, gives
 ${\bf E}_{\mbox{\footnotesize{reg}}}(0)=\mp(2e/3)\dot {\bf a}$.
Thus the only well defined part of the electric field at the location
of the particle in its instantaneous rest frame,
 ${\bf E}_{\mbox{\footnotesize{reg}}}(0)$,
 has some definite value which is ultimately determined (via the Maxwell
equations) by the fields, particles and boundary conditions elsewhere.

We are now set to establish our general conclusion:  that accelerated
charges constantly emit and absorb power at the Larmor rate.  The key
is the singular expansion of the field near the charge.  Working in
the reference frame in which the charge is instantaneously at rest at
$t=0$, we calculate (to first order is sufficient)
 the Poynting vector: ${\bf S}=(1/4\pi){\bf E} \times {\bf B}$,
using
${\bf E}(t,{\bf r})={\bf E}(0,{\bf r})+\dot {\bf E}(0,{\bf r})t$,
 along with
$\dot{\bf E}(0,{\bf r})= \nabla\times {\bf B}(0,{\bf r})$,
and similar equations for ${\bf B}(t,{\bf r})$.
The power is determined by the flux
integral of ${\bf S}\cdot \hat {\bf r}$. The contribution
 in the small $r$ limit is just the $O(1/r^2)$ part.
 To sufficient accuracy,
\begin{eqnarray}
  r^2 {\hat {\bf r}}\cdot {\bf E}\times{\bf B}&\simeq&
r^2 {\hat {\bf r}}\cdot
\left({\bf E}_{-2} + {\bf E}_{-1}+{\bf E}_0
+{\bf \nabla}\times{\bf B}_0 t\right)  
\times
\left\{{\bf B}_0-{\bf \nabla}\times
({\bf E}_{-2}+{\bf E}_{-1}+{\bf E}_0)t\right\}\nonumber\\
&\equiv& r^2 {\hat {\bf r}}\cdot
\left({\bf E}_{-1}+{\bf E}_0
+{\bf \nabla}\times{\bf B}_0 t\right)  
\times
\left\{{\bf B}_0 -{\bf \nabla}\times
({\bf E}_{-1}+{\bf E}_0)t\right\}.
\end{eqnarray}
Dropping terms which vanish in the $t^2=r^2\to 0$ limit gives
\begin{eqnarray}
r^2 {\hat {\bf r}}\cdot {\bf E}\times{\bf B}&\simeq&
r^2 {\hat {\bf r}}\cdot {\bf E}_{-1} \times
\left\{-{\bf \nabla}\times {\bf E}_{-1} t\right\} \nonumber\\
&=&
 r^2 {\hat {\bf r}}\cdot\left(- e{{\bf a}\over 2r}\right)\times
\left\{ e{{\bf a}\times{\hat{\bf r}}\over r^2}\right\} t\nonumber\\
&=&
e^2 \left|{\hat {\bf r}}\times {\bf a}\right|^2(t/ r)
=e^2 a^2 \sin^2\theta(t/r).
\end{eqnarray}
Then integration over angles gives $(2/3)e^2 a^2 (t/r)$,
the Larmor power rate multiplied by $t/r$.  This factor is $+1$
for the outgoing future null cone, the usual emission {\em but}, for
$t<0$, the factor is $-1$, indicating absorption of power
incoming along the past null cone.  Doing the calculation in this
fashion reveals that the result actually depends only on ${\bf
E}_{-1}$, the $a/r$ part of the electric field, which is
independent of the choice of advanced or retarded potential.  Now that
the physics is clear, we could do a rigourous covariant
calculation.  The details need not be given here, it is sufficient to
take Rohrlich's establishment of the {\it local criterion}
\cite{Roh61} for the emitted power and time reverse it, replacing
retarded by advanced quantities, to find {\it exactly} the same
rate for the absorbed power.

Thus an accelerated charge always emits and absorbs at the Larmor rate
locally.  There is a continuous stream of power at the Larmor rate
through the location of the charge, not a flow from the charge to the
field.  The charge seems merely to focus the ambient field flow.  Such
a vision undermines the usual argument for a radiation reaction force
based on irreversible emission.

How then should a charge move?  Our detailed considerations will be
presented elsewhere.  Briefly, we expect that a point charge would
move according to some suitably adjusted version of the Lorentz force
law.  The ``physics is simple when analyzed locally'' philosophy is
most easily applied in the instantaneous rest frame of the charge.
There the Lorentz force law reduces to the form ${\bf
F}_{\mbox{\footnotesize{em}}}=e{\bf E}$.  The complication is that
${\bf E}$ includes a singular part due to the charge's self field,
which does not have a well defined value at the location of the
charge.  The simplest, most obvious assumption is just to remove these
singular terms \cite{noself}.  Hence we propose that an
instantaneously stationary point charged particle interacts with the
only well defined part of the electric field at the charge location:
${\bf E}_{\mbox{\footnotesize{reg}}}:={\bf E}-
{\bf E}_{\mbox{\footnotesize{sing}}}$.
(Indeed, is there any other option that makes sense?)\ \
This {\em renormalized Lorentz force}
\begin{equation} {\bf F}_{\mbox{\footnotesize{em}}}= e{\bf
E}_{\mbox{\footnotesize{reg}}}(0),
\end{equation}
has, in effect, already been proposed by Kijowski \cite{Kij94}.
He took a different approach proposing certain
boundary conditions on the field at the location of the particle.  He showed
that the Maxwell equations then preserved a suitably ``renormalized'' (by
extracting the infinite Coulomb energy) expression for the total
energy-momentum of the particle-electromagnetic field system.  More
recently the initial value problem for this ``renormalized
electrodynamics'' for point charged particles has been considered
\cite{GKZ97}.  It was found that unique solutions exist, there are no
runaway solutions.


This work was supported by the National Science Council
of the R.O.C. under grant Nos.
NSC88-2112-M-008-013 and NSC88-2112-M-008-018.



\begin{references}
\bibitem{noself}
This assumption resembles the direct action electrodynamics, where
singularity problems are avoided because a particle does not interact
with itself.  However here we have only removed the singular part of
the self field.
\end{references}
\end{document}